\documentclass[twocolumn]{aastex7}
\usepackage{CJK}

\usepackage{multirow}
\begin{document}
\begin{CJK*}{UTF8}{gbsn}

\title{The anomalously large transiting planet around Boyajian's star: evidence for a circumplanetary ring}

\author[orcid=0000-0003-4027-4711]{Wei Zhu (祝伟)}
\affiliation{Department of Astronomy, Tsinghua University, Beijing 100084, China}
\email[show]{weizhu@tsinghua.edu.cn}  

\author[orcid=0009-0007-6412-0545]{Quanyi Liu (刘权毅)}
\affiliation{Department of Astronomy, Tsinghua University, Beijing 100084, China}
\email[]{liu-qy24@mails.tsinghua.edu.cn}

\correspondingauthor{Wei Zhu}

% \author[orcid=0000-0000-0000-0001,sname='North America']{Tundra North America}
% \altaffiliation{Kitt Peak National Observatory}
% \affiliation{University of Saskatchewan}
% \email[show]{fakeemail1@google.com}  

\begin{abstract}
A recent study reported a transiting giant-planet candidate with a mass of $9.4^{+4.9}_{-4.1}\,M_{\rm J}$ and a radius of $1.7\pm0.1\,R_{\rm J}$ on a wide orbit around Boyajian's star (KIC~8462852, also called Tabby's star), which is well known for its irregular, deep flux dips. The reported radius is puzzling, because cool giant planets of $\sim$10\,$M_{\rm Jup}$ are expected to have radii close to Jupiter's, so the companion appears $\sim$60--70\% larger than standard models allow. Here we explore the solution that the planet is actually ringed. We develop a fast and accurate code to compute the transit light curve due to a ringed planet, and apply it to the TESS light curve. The ring model is marginally favoured by the data by $\Delta \chi^2>10$ and a difference in Bayesian information criterion (BIC) of $<2$. The preferred configuration is a Jupiter-sized ($1.1\,R_{\rm J}$) planet surrounded by an opaque, highly flattened ring extending to $3.1$--$4.4$ planetary radii with a nearly polar obliquity. While the ring geometry is uncertain, simple estimates suggest that a debris annulus extending out to the planet's Roche limit is needed to explain the large transit depth. The ring thus removes the radius anomaly altogether and is qualitatively consistent with the dusty, comet-shepherding environment inferred from the star's historic dimmings. 
\end{abstract}



\section{Introduction}           %% first-level sections will be auto-capitalized
\label{sect:intro}

KIC~8462852, better known as Boyajian's star or Tabby's star, is an otherwise ordinary F3\,V main-sequence star that has become one of the most enigmatic objects in the \emph{Kepler} field. During the four years of \emph{Kepler} monitoring it exhibited a series of abrupt, aperiodic flux dips, reaching depths of 15\% and 22\% in events near days 792 and 1519 of the mission, which were far deeper than any planetary transit and had no detectable periodicity \citep{Boyajian2016}. 
A series of dipping events have also been observed after the \emph{Kepler} era \citep[e.g.,][]{Boyajian:2018, Bourne:2018}.
The dips are chromatic, with stronger attenuation at shorter wavelengths, indicating occulting material of dust-sized grains rather than a solid body \citep{Deeg:2018}. A wide range of mechanisms has been proposed to explain the variability of Boyajian's star, including disintegrating exocomets or comet swarms \citep{Boyajian2016,Kiefer2017}, a family of Trojan bodies sharing the orbit of a massive perturber \citep{Ballesteros2018}, debris from a recently consumed planet \citep{Metzger2017}, a warped or clumpy circumstellar dust disk, and, after all, non-natural ``megastructure'' scenarios \citep{Wright2016}. None of these has gained universal acceptance, and the leading class of models still requires an unseen massive companion whose gravity shepherds the dust-producing bodies onto star-grazing orbits. A stellar companion to Boyajian's star was discovered later \citep{Pearce2021}, but with a separation $\sim880\,$au it is unlikely to be responsible for the mysterious light curve variation of Boyajian's star.

Recently, \citet{MadurgaFavieres2026} reported a single, isolated transit event in the Transiting Exoplanet Survey Satellite \citep[TESS,][]{Ricker2014} photometry. The event is strikingly different from the irregular dips that have been seen before: it is symmetric to within the photometric precision, with a depth of $\sim$1.1\% and a duration of $\sim$21 hours, exactly the morphology expected for a solid, roughly spherical body crossing the stellar disk. Combining the transit data with a re-analysis of archival radial velocities, \citet{MadurgaFavieres2026} obtained a companion mass of $9.4^{+4.9}_{-4.1}\,M_{\rm J}$ and a radius of $1.7\pm0.1\,R_{\rm J}$ on an orbit with a period of at least $\sim$1030~days, with only five discrete period windows surviving the \emph{Kepler} archival constraints. If confirmed, this companion would both constitute the first transiting giant planet/brown dwarf around Boyajian's star and may be the long-sought perturber capable of driving the exocomet population responsible for the historic dimmings.

\begin{figure}
    \centering
    \includegraphics[width=\linewidth]{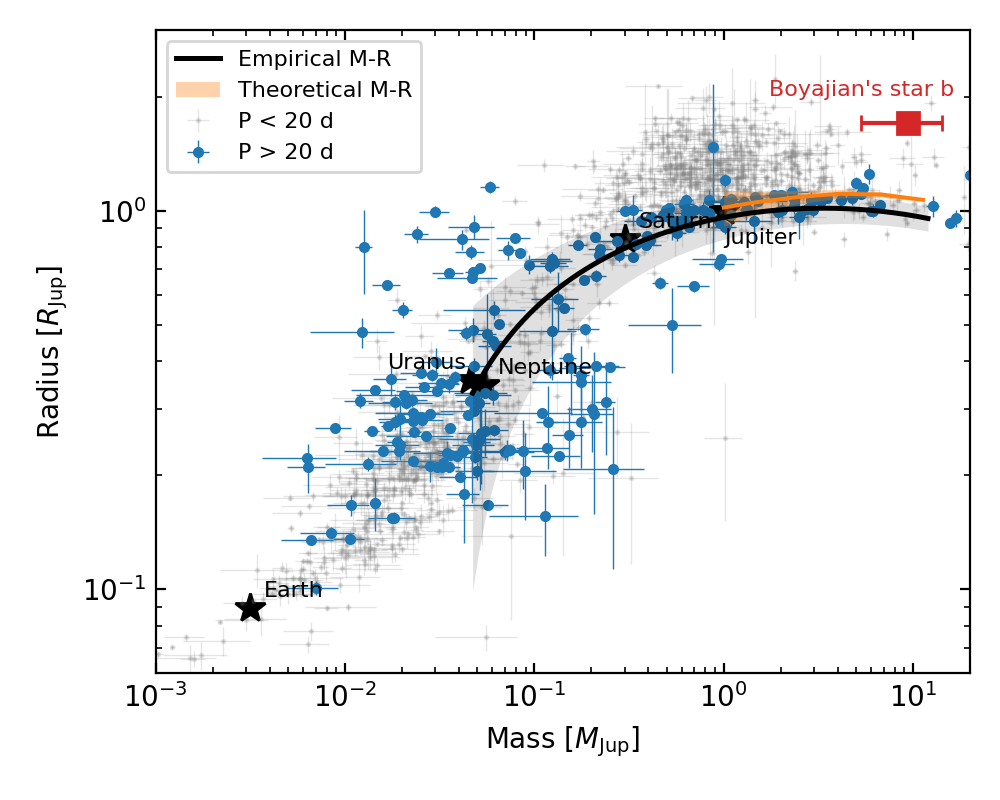}
\caption{Mass--radius diagram of known transiting exoplanets with mass and radius measured to better than 50\% uncertainty, taken from the NASA Exoplanet Archive \citep{Christiansen2025}. Planets with orbital periods $P<20$~d and $P>20$~d are shown in gray and green, respectively, and Solar System planets are also marked. The black curve with the gray band is the empirical mass--radius relation for cool ($T_{\rm eq}<1000$~K) giants \citep{Thorngren2019}, and the orange band shows the theoretical relation of giant planets of 1~Gyr at 1~au from \citet{Fortney:2007}. The recently reported planetary companion to Boyajian's star by \citet{MadurgaFavieres2026}, denoted here Boyajian's star~b, is shown as red square. Its radius is $\sim$60--70\% larger than what is expected for such a cool, massive planet.}
    \label{fig:m-r-diagram}
\end{figure}

There is, however, a physical puzzle hidden in the numbers. Gas-giant planets more massive than Jupiter are supported largely by electron degeneracy pressure, and their radii are expected to be nearly mass-independent at cool temperatures. As shown in Figure~\ref{fig:m-r-diagram}, both planetary evolution models \citep[e.g.,][]{Fortney:2007} and the empirical mass-radius relation for cool giants \citep{Thorngren2019} predict $R_{\rm p} \simeq 1.0$--$1.1\,R_{\rm J}$ across $\sim$2--12\,$M_{\rm J}$, with an intrinsic dispersion of only $\sim$0.1\,$R_{\rm J}$. Inflation by stellar irradiation, which applies to close-in hot Jupiters, is negligible at the companion's equilibrium temperature of $\sim$270~K. At an age of $\sim1.2\,$Gyr \citep{Pearce2021}, the star is no longer young for the residual heat of formation to keep the planet substantially puffed up either. The reported radius of $1.7\,R_{\rm Jup}$ is therefore surprisingly large by $\sim$60--70\%.
%A radius this large would make the object a genuine outlier in the cool-giant population and, ironically, would require an explanation of the same kind as the historic dimmings of its own host star.

In this paper we propose and quantify a geometric solution to the puzzle by adding a ring to the planetary companion. An opaque ring surrounding a much smaller planet blocks additional starlight, mimicking an anomalously large planetary radius. Similar ideas have been proposed to explain other planets. In particular, \citet{Piro2020} showed that several of the lowest-density ``super-puff'' planets could in principle be explained by small rocky planets with opaque rings, and ringed configurations have also been invoked for individual ambiguous systems \citep[e.g.][]{Aizawa2017}. In the case of Boyajian's star, a qualitatively similar idea was advanced by \citet{Ballesteros2018}, who interpreted the historic, asymmetric 15\% dip at D792 as the grazing transit of a Saturn-like ringed giant, with the later dimmings caused by a swarm of Trojan bodies sharing its orbit. The ring model in \citet{Ballesteros2018} is only schematic, because no ring parameters were fitted, the light-curve reproduction was qualitative, and the predicted repeating events, with Trojan passages in 2021 and a new transit in 2023, have not been reported. 

Our work differs from \citet{Ballesteros2018} in several essential ways. First, the motivation of our ring model is to explain the abnormally large radius of the recently found transiting object, rather than to explain the overall light curve variations. As such, we aim to explain an actual, unambiguous transit event of the companion, whose orbital period and mass have been independently anchored by radial velocities \citep{MadurgaFavieres2026}. Unlike in \citet{Ballesteros2018}, we combine detailed light curve modeling with physical constraints to reveal the ring geometry. As shown later in the paper, our reported ring structure is also quite different from that proposed in \citet{Ballesteros2018}.

This paper is organized as follows. Section~\ref{sec:data} describes the TESS light curve and our standard transit fit; Section~\ref{sec:jojo} describes the new code for efficient computation of the transit light curve by ringed planet, building on the core algorithm of \citet{Liu2025}; in Section~\ref{sec:ring}, we present the ring model fit to the TESS light curve; finally, we brief discuss our findings in Section~\ref{sec:discussion}.

\section{TESS data and standard transit fit}
\label{sec:data}

\subsection{TESS light curve}

We analyzed the \emph{TESS} light curve of KIC~8462852 (Boyajian's star) obtained during Sector 15 (2019 August 15 -- September 10), which covers the single-transit event reported by \citet{MadurgaFavieres2026}. We used the pre-search data conditioned (PDC) light curve at the short-cadence (2-min) sampling produced by the Science Processing Operations Center (SPOC) pipeline \citep{Jenkins:2016}, retrieved from the Mikulski Archive for Space Telescopes (MAST). All times are given in Barycentric TESS Julian Date (BTJD $=$ BJD $-$ 2{,}457{,}000). All light-curve processing was performed with custom \texttt{python} routines built on \texttt{lightkurve} \citep{Lightkurve:2018}. 

At the given transit event identified by \citet{MadurgaFavieres2026} on 2019 September 3--4, 
%we re-derived the local baseline with a windowed polynomial detrending optimized to preserve the event shape: 
we cut a $\pm1.5$~d segment centered on the transit midpoint, masked the transit with a full window of 1.0~d, and fitted a quadratic polynomial to the out-of-transit points only. The entire segment, including transit event, was divided by this polynomial. The final product is a detrended, normalized light curve with the local baseline flat at unity and the full ingress/egress profile preserved, sampled at the native 2-min cadence, with time given in days relative to the adopted transit center (BTJD $\approx 1730.147$). The standard deviation of the normalized out-of-transit light curve is taken as the flux uncertainties.

\begin{figure*}
    \centering
    \includegraphics[width=0.9\linewidth]{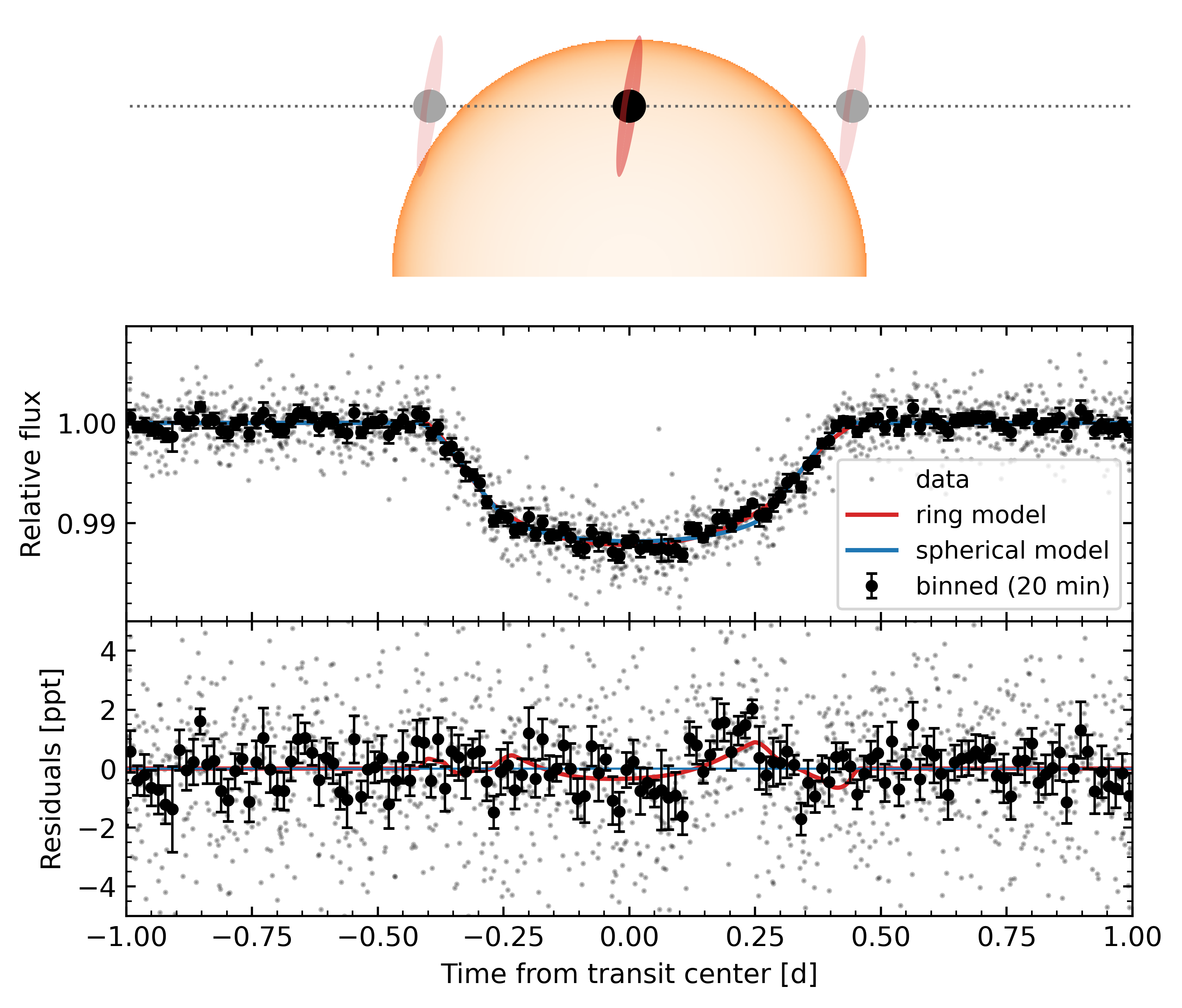}
    \caption{Best-fit standard transit (i.e., spherical planet) and ring models to the \emph{TESS} Sector~15 transit of Boyajian's star. 
    \emph{Top:} sky-projected geometry of the best-fit ring model ($r_{\rm in}=1.1\,R_{\rm p}$ case), with the upper hemisphere of the limb-darkened stellar disk shown. The panel aligns vertically with the light-curve panels below (mid-transit at $t=0$, ingress and egress at their respective epochs). The opaque system marks mid-transit, and the transparent systems mark ingress and egress (first and last contact of the model light curve). 
    \emph{Middle:} detrended 2-min \emph{TESS} photometry (grey points; binned to 20~min with $1\sigma$ error bars) with the best-fit ring model (red; $r_{\rm in}=0.077$, $r_{\rm out}=0.300$, $f_{\rm r}=0.89$, $\theta=82^\circ$) and the best-fit spherical model (blue; $R_{\rm p}/R_\star=0.109$, $b=0.77$) overplotted. \emph{Bottom:} residuals of the spherical model and the difference between the ring and spherical model light curves (red), in parts per thousand (ppt). The ring redistributes occulted area to absorb the structured residuals near the transit bottom.}
    \label{fig:lcs}
\end{figure*}

\begin{table*}
  \centering
  \caption{Best-fit parameters of the transit models for the TESS Sector~15 transit event of Boyajian's star. Median values with the 16th and 84th percentiles are given for fitted parameters. The standard quadratic limb darkening parameters, $u_1$ and $u_2$, are derived from the $q_1$ and $q_2$ parameters of \citet{Kipping:2013}. The reference epoch of the detrended light curve is BTJD $=1730.147$.}
  \label{tab:fits}
  \begin{tabular}{lccc}
  \hline\hline
  Parameter & Spherical & Ring ($r_{\rm in}=R_{\rm p}$) & Ring ($r_{\rm in}=1.1R_{\rm p}$) \\
  \hline
  Mid-transit time $t_{\rm c}$ [d] & $0.0084\pm{0.0026}$ & $0.0060\pm{0.0028}$ & $0.0054\pm{0.0029}$ \\
  Impact parameter $b$ & $0.77\pm0.03$ & $0.79^{+0.04}_{-0.03}$ & $0.72\pm{0.04}$ \\
  Planet radius $R_{\rm p}/R_\star$ & $0.1090^{+0.0016}_{-0.0017}$ & 0.07 (fixed) & 0.07 (fixed) \\
  Ring inner radius $r_{\rm in}/R_\star$ & -- & 0.07 (fixed) & 0.077 (fixed) \\
  Ring outer radius $r_{\rm out}/R_\star$ & -- & $0.22^{+0.04}_{-0.06}$ & $0.31^{+0.06}_{-0.07}$ \\
  Ring flattening $f_{\rm r}$ & -- & $0.82^{+0.13}_{-0.08}$ & $0.89^{+0.07}_{-0.04}$ \\
  Ring obliquity $\theta$ [deg] & -- & $74^{+10}_{-7}$ & $81^{+5}_{-4}$ \\
  $\log_{10}\rho_\star$ [g cm$^{-3}$] & $-0.30\pm{0.06}$ & $-0.31\pm{0.09}$ & $-0.12^{+0.07}_{-0.06}$ \\
  $q_1$ & $0.44\pm{0.09}$ & $0.31^{+0.09}_{-0.10}$ & $0.27\pm{0.10}$ \\
  $q_2$ & $0.28\pm{0.11}$ & $0.22\pm0.10$ & $0.21\pm0.10$ \\
  $u_1$ (derived) & $0.37^{+0.15}_{-0.16}$ & $0.24^{+0.11}_{-0.12}$ & $0.21^{+0.10}_{-0.12}$ \\
  $u_2$ (derived) & $0.29\pm{0.14}$ & $0.30^{+0.11}_{-0.12}$ & $0.29^{+0.11}_{-0.12}$ \\
  Model $\chi^2$ ($\pm0.5$~d around $t_c$) & 644.7 & 631.2 & 630.4 \\
  $\Delta$ BIC & 0 & $-0.4$ & $-1.3$ \\
  \hline
  \end{tabular}
\end{table*}

\subsection{Standard transit fit} \label{sec:spherical}

For consistency check, we have performed the standard \citet{MandelAgol2002} model fit to the flattened light curve. The model parameters are the mid-transit time $t_0$, the impact parameter $b$, the planet-to-star radius ratio $R_{\rm p}/R_\star$, the quadratic limb-darkening coefficients $(q_1,q_2)$ \citep{Kipping:2013}, and the stellar density $\rho_\star$, which sets the orbital scale $a/R_\star$ at fixed period. Because a single transit does not constrain the orbital period or the orbital shape, we fixed $P = 1030$~d, $e = 0$ and $\omega = 90^\circ$.
We used uniform priors over physically motivated ranges for all free parameters, with the exception of the limb-darkening coefficients. For $q_1$ and $q_2$, we instead adopted Gaussian priors centered on the theoretical values ($0.26$ and $0.19$, respectively) from \citet{Claret:2017}, evaluated at the spectroscopic stellar parameters of Boyajian's star \citep{Boyajian2016}, with standard deviations of $0.10$. We sampled the posterior with the affine-invariant MCMC ensemble sampler \texttt{emcee} \citep{emcee2013}, and the chains were run until the integrated autocorrelation times were much shorter than the discarded burn-in. The resulting parameters and associated uncertainties are listed in Table~\ref{tab:fits}.

Our spherical-fit results are in good agreement with those of \citet{MadurgaFavieres2026} despite the very different methodology. In particular, our radius ratio agrees within $1\sigma$ ($0.1090^{+0.0017}_{-0.0016}$ here vs.\ $0.111\pm0.004$), as does the impact parameter ($0.77\pm0.03$ vs.\ $0.76^{+0.05}_{-0.04}$), and our inferred mean stellar density ($\simeq$0.50 g cm$^{-3}$) matches their joint-fit value ($0.49\pm0.10$ g cm$^{-3}$) within the errors. Consequently, the derived companion radius from our fit, $R_{\rm c}\simeq1.74\,R_{\rm Jup}$, is consistent with their $1.7\pm0.1\,R_{\rm Jup}$, confirming that it is unusually large.

The reduced $\chi^2$ of the best-fit model is close to unity, although structured residuals near the transit bottom (Figure~\ref{fig:lcs}) hint at deviations from a perfect spherical object. As will be shown in Section~\ref{sec:ring}, they can be largely explained by the ring model.

%%%%
\section{Implementation of the ring model in \texttt{JoJo}} \label{sec:jojo}

The calculation of the transit light curve due to a ringed planet is non-trivial, especially for the purpose of MCMC modeling. Here we adopt the key algorithm of \texttt{JoJo}, first proposed in \citet{Liu2025} and used to model the transit light curve due to an oblate planet \citep{Liu2024}, and apply it to the ring situation.
\footnote{\texttt{JoJo} code is publicly available at \url{https://github.com/Flippedx/JoJo}.}
Our ring model includes a spherical body and a flat, circular ring viewed at an arbitrary obliquity. The ring is defined by its inner and outer radii ($r_{\rm in}$ and $r_{\rm out}$, in units of $R_\star$), a flattening factor $f_{\rm r}$, the sky-projected obliquity $\theta_\perp$ of the ring's major axis, and an effective opacity $\alpha$ that controls the transparency of the ring. The aspect ratio of the projected elliptical shape of the ring is given by $b/a=1-f_{\rm r}$. A schematic plot of the ring model is illustrated in the upper right corner of Figure~\ref{fig:corners}.

The occultation by the spherical planet follows the quadratic limb-darkened formalism of \citet{MandelAgol2002}, while the ring contribution is obtained by integrating the stellar intensity over the overlap of the stellar disk with the elliptical annulus between the inner and outer ring edges. For each exposure the code solves the mutual intersections of the three bounding curves (stellar limb and the two ring ellipses, with the planet--ring intersections where the planet overlaps the ring hole) and evaluates the occulted flux through boundary integrals of the intensity along those arcs \citep{Liu2025}, so all topological cases---ring fully on the star, ingress/egress, partial stellar limb crossings, and planet--ring overlap---are treated without heavy numerical grid integration. Limb darkening again uses the \citet{Kipping:2013} triangular parameterization, and the orbit is parameterized as in Sect.~\ref{sec:spherical}. 

We validate our ring code against \texttt{PYPPLUSS} \citep{Rein2019} and find it more efficient. To achieve the same precision, \texttt{JoJo}'s ring module is $>20\times$ faster than \texttt{PYPPLUSS}.

\section{Ring model fit to the TESS light curve} \label{sec:ring}

Given the relatively high noise level of TESS data and the flexibility of the ring model, it is necessary to make a few reasonable assumptions about the ring model in order to make it physically plausible. Given the mass constraint on the transiting companion from \citet{MadurgaFavieres2026}, we can adopt the theoretically predicted radius according to the well understood mass--radius relation of cold giant planets (e.g., \citealt{Fortney:2007}; see also Figure~\ref{fig:m-r-diagram}), $R_{\rm p}=1.1~R_{\rm J}$, regardless of the assumed core mass. This gives $R_{\rm p}/R_\star=0.07$. 

The outer radius of the ring should be within the fluid Roche limit \citep{Murray:1999}
\begin{equation} \label{eqn:roche}
    R_{\rm r} \approx 2.46 \left(\frac{\rho_{\rm p}}{\rho_{\rm r}}\right)^{1/3} R_{\rm p} ,
\end{equation}
where $\rho_{\rm p}$ and $\rho_{\rm r}$ are the mean densities of the planet and the ring particles, respectively. At the estimated separation of $\approx 2.2~$au, the planet is well within the water ice line of the F3V host, so the ring particles should be dominated by rocky rather than icy material. While both $\rho_{\rm p}$ and $\rho_{\rm r}$ are around the same order, their values may vary by a factor of a few. For example, the inferred mean density of the planet would be $\sim9~$g~cm$^{-3}$ with a mass of $10~M_{\rm J}$ and a radius of $1.1~R_{\rm J}$, whereas the density of the ring particles can be as low as $\sim1~$g~cm$^{-3}$ \citep[e.g.,][]{Watanabe:2019, Scheeres:2019}. As so, we consider a ring outer radius $r_{\rm out}\gtrsim5~R_{\rm p}$ highly unlikely, although a more quantitative upper bound is somewhat uncertain.

The inner ring radius is also uncertain. At equilibrium, the inner boundary is set by the balance between the supply due to collisional grinding and loss due to various drags, but its location cannot be determined from simple estimations in the lack of detailed physical properties of the planet \citep[e.g.,][]{Goldreich1982, Cuzzi2010}. 
As will be shown later in the case of Boyajian's star, $r_{\rm in}$ of the ring model is not well constrained by itself, but a larger $r_{\rm in}$ pushes $r_{\rm out}$ out of the physically plausible range because the cross-sectional area should be conserved. For practical purposes, we therefore consider two quite extreme cases, $r_{\rm in}=R_{\rm p}$ and $r_{\rm in}=1.1R_{\rm p}$. This latter case matches Saturn's ring system, with the D ring extending to $\sim1.1$ times of the Saturn's radius. 

We assume a fully opaque ring with $\alpha = 1$, and fix $P$, $e$, and $\omega$ in the same way as in the standard fit. For the limb darkening parameters, the same parameterization and priors as in the standard fit are adopted.
%The ring model fit therefore take nine parameters: $(t_0, b, r_{\rm in}, r_{\rm out}, f_{\rm r}, \theta, \log_{10}\rho_\star, q_1, q_2)$. 
We again use \texttt{emcee} to sample the posterior. The final run uses 64 walkers each with 10,000 steps, and the first 3000 steps are excluded from the posterior estimation.

\begin{figure*}
\includegraphics[width=0.95\textwidth]{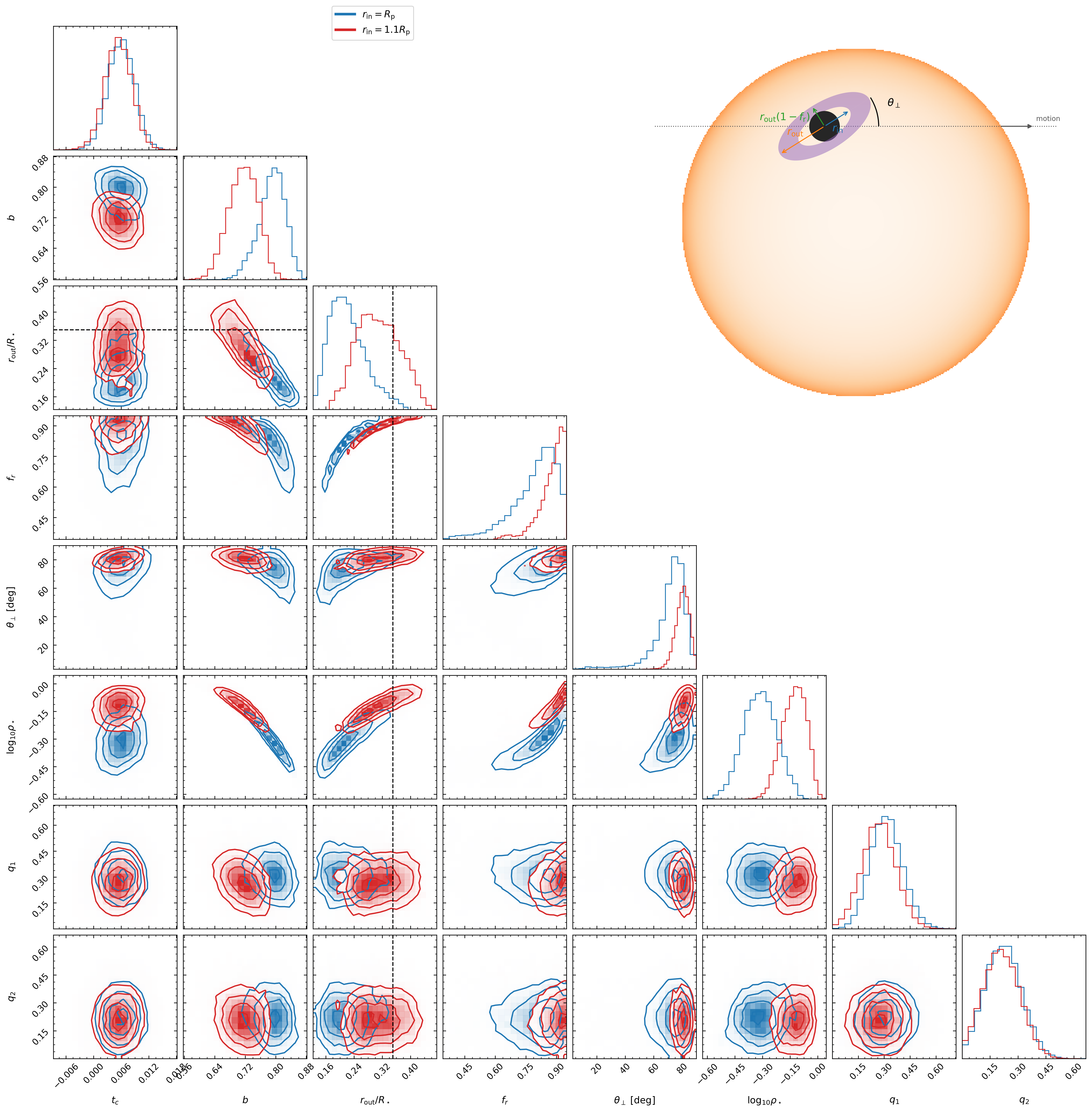}
\caption{Comparison of the ringed-planet MCMC posteriors for two assumptions on the fixed inner ring radius: $r_{\rm in} = R_{\rm p}$ (blue) and $r_{\rm in} = 1.1\,R_{\rm p}$ (red). The dashed vertical lines mark $r_{\rm out} = 5\,R_{\rm p} = 0.35 R_\star$ in every panel involving $r_{\rm out}/R_\star$. The two solutions are offset but overlapping: enlarging the inner radius moves the outer radius outward, increases the ring flatness and obliquity, and shifts the chord to smaller $b$, illustrating the degeneracy between the ring geometry and the orbital configuration. 
The inset in the upper right corner illustrates the meaning of the model parameters related to the ring. Note that this cartoon geometry does not use the best-fit values of the ring model.}
\label{fig:corners}
\end{figure*}

Results from the ring model fits with two fixed $r_{\rm in}$ are given in Table~\ref{tab:fits}, and the posteriors are also illustrated in Figure~\ref{fig:corners}. Note that the goodness-of-fit statistics ($\chi^2$ and BIC) are evaluated on the 718 data points within $\pm 0.5$~d of the fitted mid-transit time, on the basis that the out-of-transit baseline has no contribution to the transit signal. Because the two ring models are nearly identical, we choose the one with $r_{\rm in}=1.1R_{\rm P}$ and show its best-fit ring model along with the transit data and the best-fit spherical model in Figure~\ref{fig:lcs}. Compared to the standard spherical model, the ring solutions replace most of the planet's cross-section with an extended, highly flattened annulus. With the planetary radius fixed at $R_{\rm p}/R_\star = 0.070$, the ring outer radius is $r_{\rm out} \simeq 0.22$--$0.31\,R_\star$, corresponding to $\sim3.1$--$4.4\,R_{\rm p}$, and the ring has a flatness of $f_{\rm r} \simeq 0.8$--$0.9$ and a projected obliquity of $\theta_\perp \simeq 74$--$81^\circ$. Assuming a concentric and circle ring, we can infer its inclination with respect to the line of sight $i_{\rm r}=\cos^{-1}(1-f_{\rm r})\approx 78$--$84^\circ$. The above values have relatively large uncertainties, because the ring-related parameters are not well constrained by the TESS data and there is a coherent degeneracy between parameters describing the ring geometry and those describing the transit geometry (e.g., the impact parameter $b$ and the orbital scale $\log_{10}\rho_\star$), as shown in Figure~\ref{fig:corners}. Nevertheless, our ring model fits suggest that the transit event of Boyajian's star can be explained by a ring that is highly inclined with respect to the line of sight, nearly perpendicular to the transit chord, and spreading almost all the way to its Roche limit.

In terms of goodness of fit, both ring models improve the fits by $\Delta\chi^2 \simeq 14$ relative to the spherical model, at the expense of two additional free parameters. The improvement is concentrated near the transit bottom, where the spherical model leaves structured residuals (Figure~\ref{fig:lcs}). After penalizing the extra parameters, the statistical evidence for the ring model is reduced, but it is still slightly favoured over the spherical model by $\Delta{\rm BIC} \simeq -0.4$ (for $r_{\rm in}=R_{\rm p}$) and $-1.3$ (for $r_{\rm in}=1.1R_{\rm p}$). 

The ring model tries to explain the light curve feature at 0.25~d offset from the mid-transit. Such a feature may be caused by other astrophysical effects such as spot crossing, which may not be rare for a relatively young and fast-rotating F3-type dwarf \citep{Boyajian2016}. If so, what does it mean to the ring model? Excluding that light curve feature in the fits undoubtedly overturns the model comparison metric. However, because the standard transit model yields an unreasonably large size for the planetary-mass companion, the physical motivation for the introduction of the ring remains strong. From simple arguments, one can conclude that the ring, if exists, should be quite extended. Given the well constrained transit depth, the outer radius of the ring is $r_{\rm out}\geq 0.11/0.07\approx 1.6 R_{\rm p}$, with the lower limit achieved if the ring is exactly face-on and connects all the way to the planet. In the more likely scenario of an inclined ring, $r_{\rm out}\gtrsim 1.6 R_{\rm p}/\sqrt{\cos{i_{\rm r}}}$,
\footnote{The lower limit is reached if there is no overlap between the planetary disk and the projected ring.}
so a Saturn-like obliquity ($26.7^\circ$) would push the outer radius to $\gtrsim2.4R_{\rm p}$, which is already close to the Roche limit (Equation~\ref{eqn:roche}) and well consistent with the inferred values from our ring model.

Although an extended ring is required regardless of the detailed light curve morphology, its orientation is largely set by the feature at $t\sim0.25\,$d after mid-transit. With $b\approx0.8$, a highly oblique ring begins egress earlier than a spherical planet with the same cross-sectional area. Light curves of the two models begin to differ, and the difference between them therefore leaves a rebrightening feature, explaining the feature at $t\sim0.25\,$d in the data.

With a nearly polar obliquity, the derived ring geometry is nevertheless intriguing. The true obliquity of the ring can be derived from its inclination $i_{\rm r}$ and the projected obliuqity $\theta_\perp$, via $\theta = \cos^{-1}(\pm \sin i_{\rm r} \cos \theta_\perp)$, and the $\pm$ signs concern the ambiguity of the ring orientation. Adopting $80^\circ$ for both $i_{\rm r}$ and $\theta_\perp$, the true obliquity is $\sim80^\circ$ or $\sim100^\circ$. Taking it as the spin direction of the planet, this super Jupiter is then rotating sideways, very similar to Uranus of our Solar System.

\section{Discussion} \label{sec:discussion}

We have presented a quantitative comparison of two interpretations of the single TESS Sector~15 transit of Boyajian's star reported by \citet{MadurgaFavieres2026}: a bare spherical planet, and a Jupiter-sized planet surrounded by an opaque ring. Both models reproduce the light curve at formally acceptable levels, and the ring improves the fit by $\Delta\chi^2 \simeq 14$ for two additional parameters, concentrated at the structured residual near the transit bottom (Figure~\ref{fig:lcs}). After the penalty due to two more free parameters is taken into account, the two interpretations are statistically nearly indistinguishable, with $\Delta{\rm BIC} \simeq -0.4$ to $-1.3$ in favour of the ring model. 

Although the statistical evidence for the ring is fairly marginal, the physical motivation for it is strong. The spherical fit requires a planet of $R_{\rm p} \simeq 1.7\,R_{\rm Jup}$ at $M_{\rm p} \simeq 9.4\,M_{\rm Jup}$, which is $\sim$60--70\% larger than the radii predicted by theoretical models and the empirical mass--radius relation (Figure~\ref{fig:m-r-diagram}). The ring fit absorbs most of the occulted area into a highly flattened, nearly polar annulus extending to $3.1$--$4.4$ planetary radii, and thus the planetary radius is no longer abnormal.

If the ring is real, it has several implications for our understanding of the system as well as the ring formation in general. First, the ring extends nearly to the companion's Roche limit. Given the location of the companion, it is a debris ring likely sustained by the disruption of comets or small satellites, which may be the same minor-body reservoir invoked to explain the historic dipping activity of this star. The ring component may also explain the brightening seen in the TESS data within $\sim5$~d before and after the transit, which \citet{MadurgaFavieres2026} also attribute to forward-scattering by possible circumplanetary dust. A detailed modeling of this feature in the context of our ring model is beyond the scope of the present work. If ringed, the companion would then simultaneously be the perturber that shepherds exocomets onto star-grazing orbits and a dusty object in its own right, making the system a scaled-up analogue of a ringed giant plus cometary swarm. 

The nearly polar ring obliquity ($\theta \simeq 80^\circ$ or $100^\circ$) implies that the ring plane is strongly tilted with respect to the orbital plane of the companion, a configuration reminiscent of Uranus and plausibly produced by the tidal disruption of a captured satellite or by obliquity excitation during the system's dynamical history. We caution, however, that the ring orientation is only weakly constrained by the single available transit of relatively poor data quality, and that the apparent feature that drives the ring fit could in part be spot crossing (Sect.~\ref{sec:ring}).

The ring and no ring interpretations make sharply different predictions for observations that directly probe the companion. The very direct test is precise photometry of the repeat transits, with the next expected in the 2026 October and 2027 January/March windows \citep{MadurgaFavieres2026}. The ring and spherical models differ most at ingress and egress, where the ring extends the duration of the partial phase and distorts its shape, and the predicted difference is at the few$\times10^{-4}$ level, which is within reach of space-based time-series photometry of existing facilities, although the currently uncertain transit ephemerides makes the scheduling difficult. Depending on the composition and opacity of the ring particles, simultaneous multi-band photometry may also be able to test the ring model against the spherical planet model.
Finally, with the upcoming Gaia~DR4 epoch astrometry, the orbital properties and mass of the companion can be independently constrained \citep{MadurgaFavieres2026}. These measurements may provide conclusive evidence for the radius anomaly, even though they may not be able to test directly the ring model.

%% Please use the acknowledgment and contribution environments. This will 
%% be anonomyized when the "anonymous" style option is used. 
\begin{acknowledgments}
Work by W.Z.\ and Q.L.\ was funded by the National Natural Science Foundation of China (NSFC) under No.\ 12133005.
This paper includes data collected by the TESS mission, which are publicly available from the Mikulski Archive for Space Telescopes (MAST). Funding for the TESS mission is provided by NASA's Science Mission directorate. 
\end{acknowledgments}

\bibliography{bibtex}{}

@ARTICLE{Claret:2017,
       author = {{Claret}, A.},
        title = "{Limb and gravity-darkening coefficients for the TESS satellite at several metallicities, surface gravities, and microturbulent velocities}",
      journal = {\aap},
         year = 2017,
        month = apr,
       volume = {600},
          eid = {A30},
        pages = {A30},
          doi = {10.1051/0004-6361/201629705},
archivePrefix = {arXiv},
       eprint = {1804.10295},
 primaryClass = {astro-ph.SR},
       adsurl = {https://ui.adsabs.harvard.edu/abs/2017A&A...600A..30C}
}

@ARTICLE{Kipping:2013,
       author = {{Kipping}, David M.},
        title = "{Efficient, uninformative sampling of limb darkening coefficients for two-parameter laws}",
      journal = {\mnras},
         year = 2013,
        month = nov,
       volume = {435},
       number = {3},
        pages = {2152-2160},
          doi = {10.1093/mnras/stt1435},
archivePrefix = {arXiv},
       eprint = {1308.0009},
 primaryClass = {astro-ph.SR},
       adsurl = {https://ui.adsabs.harvard.edu/abs/2013MNRAS.435.2152K}
}

@ARTICLE{Boyajian2016,
       author = {{Boyajian}, T.~S. and {LaCourse}, D.~M. and {Rappaport}, S.~A. and {Fabrycky}, D. and {Fischer}, D.~A. and {Gandolfi}, D. and {Kennedy}, G.~M. and {Korhonen}, H. and {Liu}, M.~C. and {Moor}, A. and {Olah}, K. and {Vida}, K. and {Wyatt}, M.~C. and {Best}, W.~M.~J. and {Brewer}, J. and {Ciesla}, F. and {Cs{\'a}k}, B. and {Deeg}, H.~J. and {Dupuy}, T.~J. and {Handler}, G. and {Heng}, K. and {Howell}, S.~B. and {Ishikawa}, S.~T. and {Kov{\'a}cs}, J. and {Kozakis}, T. and {Kriskovics}, L. and {Lehtinen}, J. and {Lintott}, C. and {Lynn}, S. and {Nespral}, D. and {Nikbakhsh}, S. and {Schawinski}, K. and {Schmitt}, J.~R. and {Smith}, A.~M. and {Szabo}, Gy. and {Szabo}, R. and {Viuho}, J. and {Wang}, J. and {Weiksnar}, A. and {Bosch}, M. and {Connors}, J.~L. and {Goodman}, S. and {Green}, G. and {Hoekstra}, A.~J. and {Jebson}, T. and {Jek}, K.~J. and {Omohundro}, M.~R. and {Schwengeler}, H.~M. and {Szewczyk}, A.},
        title = "{Planet Hunters IX. KIC 8462852 - where's the flux?}",
      journal = {\mnras},
         year = 2016,
        month = apr,
       volume = {457},
       number = {4},
        pages = {3988-4004},
          doi = {10.1093/mnras/stw218},
archivePrefix = {arXiv},
       eprint = {1509.03622},
 primaryClass = {astro-ph.SR},
       adsurl = {https://ui.adsabs.harvard.edu/abs/2016MNRAS.457.3988B}
}

@ARTICLE{MadurgaFavieres2026,
       author = {{Madurga-Favieres}, Cristina and {Str{\o}m}, P.~A. and {Wilson}, Thomas G. and {Heidari}, Neda and {Kiefer}, Flavien},
        title = "{Evidence for a giant companion orbiting Tabby's star}",
      journal = {\mnras},
         year = 2026,
        month = aug,
       volume = {550},
       number = {3},
          eid = {stag1273},
        pages = {stag1273},
          doi = {10.1093/mnras/stag1273},
archivePrefix = {arXiv},
       eprint = {2607.20643},
 primaryClass = {astro-ph.EP},
       adsurl = {https://ui.adsabs.harvard.edu/abs/2026MNRAS.550g1273M}
}

@INPROCEEDINGS{Jenkins:2016,
       author = {{Jenkins}, Jon M. and {Twicken}, Joseph D. and {McCauliff}, Sean and {Campbell}, Jennifer and {Sanderfer}, Dwight and {Lung}, David and {Mansouri-Samani}, Masoud and {Girouard}, Forrest and {Tenenbaum}, Peter and {Klaus}, Todd and et al.},
        title = "{The TESS science processing operations center}",
    booktitle = {Software and Cyberinfrastructure for Astronomy IV},
         year = 2016,
       editor = {{Chiozzi}, Gianluca and {Guzman}, Juan C.},
       series = {Society of Photo-Optical Instrumentation Engineers (SPIE) Conference Series},
       volume = {9913},
        month = aug,
          eid = {99133E},
        pages = {99133E},
          doi = {10.1117/12.2233418},
       adsurl = {https://ui.adsabs.harvard.edu/abs/2016SPIE.9913E..3EJ}
}

@software{Lightkurve:2018,
       author = {{Lightkurve Collaboration} and {Cardoso}, Jos{\'e} Vin{\'\i}cius de Miranda and {Hedges}, Christina and {Gully-Santiago}, Michael and {Saunders}, Nicholas and {Cody}, Ann Marie and {Barclay}, Thomas and {Hall}, Oliver and {Sagear}, Sheila and {Turtelboom}, Emma and {Zhang}, Johnny and {Tzanidakis}, Andy and {Mighell}, Ken and {Coughlin}, Jeff and {Bell}, Keaton and {Berta-Thompson}, Zach and {Williams}, Peter and {Dotson}, Jessie and {Barentsen}, Geert},
        title = "{Lightkurve: Kepler and TESS time series analysis in Python}",
 howpublished = {Astrophysics Source Code Library, record ascl:1812.013},
         year = 2018,
        month = dec,
          eid = {ascl:1812.013},
archivePrefix = {ascl},
       eprint = {1812.013},
       adsurl = {https://ui.adsabs.harvard.edu/abs/2018ascl.soft12013L}
}

@ARTICLE{MandelAgol2002,
       author = {{Mandel}, Kaisey and {Agol}, Eric},
        title = "{Analytic Light Curves for Planetary Transit Searches}",
      journal = {\apjl},
         year = 2002,
        month = dec,
       volume = {580},
       number = {2},
        pages = {L171-L175},
          doi = {10.1086/345520},
archivePrefix = {arXiv},
       eprint = {astro-ph/0210099},
 primaryClass = {astro-ph},
       adsurl = {https://ui.adsabs.harvard.edu/abs/2002ApJ...580L.171M}
}

@ARTICLE{emcee2013,
       author = {{Foreman-Mackey}, Daniel and {Hogg}, David W. and {Lang}, Dustin and {Goodman}, Jonathan},
        title = "{emcee: The MCMC Hammer}",
      journal = {\pasp},
         year = 2013,
        month = mar,
       volume = {125},
       number = {925},
        pages = {306},
          doi = {10.1086/670067},
archivePrefix = {arXiv},
       eprint = {1202.3665},
 primaryClass = {astro-ph.IM},
       adsurl = {https://ui.adsabs.harvard.edu/abs/2013PASP..125..306F}
}

@ARTICLE{Liu2025,
       author = {{Liu}, Quanyi and {Zhu}, Wei and {Zhou}, Yifan and {Hu}, Zhecheng and {Lin}, Zitao and {Dai}, Fei and {Masuda}, Kento and {Wang}, Sharon X.},
        title = "{Detecting Planetary Oblateness in the Era of JWST: A Case Study of Kepler-167e}",
      journal = {\aj},
         year = 2025,
        month = feb,
       volume = {169},
       number = {2},
          eid = {79},
        pages = {79},
          doi = {10.3847/1538-3881/ad9b8c},
archivePrefix = {arXiv},
       eprint = {2406.11644},
 primaryClass = {astro-ph.EP},
       adsurl = {https://ui.adsabs.harvard.edu/abs/2025AJ....169...79L}
}

@ARTICLE{Liu2024,
       author = {{Liu}, Quanyi and {Zhu}, Wei and {Masuda}, Kento and {Libby-Roberts}, Jessica E. and {Bello-Arufe}, Aaron and {Ca{\~n}as}, Caleb I.},
        title = "{An Extremely Low-density Exoplanet Spins Slow}",
      journal = {\apjl},
         year = 2024,
        month = nov,
       volume = {976},
       number = {1},
          eid = {L14},
        pages = {L14},
          doi = {10.3847/2041-8213/ad8f39},
archivePrefix = {arXiv},
       eprint = {2410.07977},
 primaryClass = {astro-ph.EP},
       adsurl = {https://ui.adsabs.harvard.edu/abs/2024ApJ...976L..14L}
}

@ARTICLE{Fortney:2007,
       author = {{Fortney}, J.~J. and {Marley}, M.~S. and {Barnes}, J.~W.},
        title = "{Planetary Radii across Five Orders of Magnitude in Mass and Stellar Insolation: Application to Transits}",
      journal = {\apj},
         year = 2007,
        month = apr,
       volume = {659},
       number = {2},
        pages = {1661-1672},
          doi = {10.1086/512120},
archivePrefix = {arXiv},
       eprint = {astro-ph/0612671},
 primaryClass = {astro-ph},
       adsurl = {https://ui.adsabs.harvard.edu/abs/2007ApJ...659.1661F}
}

@BOOK{Murray:1999,
       author = {{Murray}, Carl D. and {Dermott}, Stanley F.},
        title = "{Solar System Dynamics}",
         year = 1999,
          doi = {10.1017/CBO9781139174817},
       adsurl = {https://ui.adsabs.harvard.edu/abs/1999ssd..book.....M}
}

@ARTICLE{Watanabe:2019,
       author = {{Watanabe}, S. and {Hirabayashi}, M. and {Hirata}, N. and {Hirata}, Na. and {Noguchi}, R. and {Shimaki}, Y. and {Ikeda}, H. and {Tatsumi}, E. and {Yoshikawa}, M. and {Kikuchi}, S. and {Yabuta}, H. and {Nakamura}, T. and {Tachibana}, S. and {Ishihara}, Y. and {Morota}, T. and {Kitazato}, K. and {Sakatani}, N. and {Matsumoto}, K. and {Wada}, K. and {Senshu}, H. and {Honda}, C. and {Michikami}, T. and {Takeuchi}, H. and {Kouyama}, T. and {Honda}, R. and {Kameda}, S. and {Fuse}, T. and {Miyamoto}, H. and {Komatsu}, G. and {Sugita}, S. and {Okada}, T. and {Namiki}, N. and {Arakawa}, M. and {Ishiguro}, M. and {Abe}, M. and {Gaskell}, R. and {Palmer}, E. and {Barnouin}, O.~S. and {Michel}, P. and {French}, A.~S. and {McMahon}, J.~W. and {Scheeres}, D.~J. and {Abell}, P.~A. and {Yamamoto}, Y. and {Tanaka}, S. and {Shirai}, K. and {Matsuoka}, M. and {Yamada}, M. and {Yokota}, Y. and {Suzuki}, H. and {Yoshioka}, K. and {Cho}, Y. and {Tanaka}, S. and {Nishikawa}, N. and {Sugiyama}, T. and {Kikuchi}, H. and {Hemmi}, R. and {Yamaguchi}, T. and {Ogawa}, N. and {Ono}, G. and {Mimasu}, Y. and {Yoshikawa}, K. and {Takahashi}, T. and {Takei}, Y. and {Fujii}, A. and {Hirose}, C. and {Iwata}, T. and {Hayakawa}, M. and {Hosoda}, S. and {Mori}, O. and {Sawada}, H. and {Shimada}, T. and {Soldini}, S. and {Yano}, H. and {Tsukizaki}, R. and {Ozaki}, M. and {Iijima}, Y. and {Ogawa}, K. and {Fujimoto}, M. and {Ho}, T.-M. and {Moussi}, A. and {Jaumann}, R. and {Bibring}, J.-P. and {Krause}, C. and {Terui}, F. and {Saiki}, T. and {Nakazawa}, S. and {Tsuda}, Y.},
        title = "{Hayabusa2 arrives at the carbonaceous asteroid 162173 Ryugu{\textemdash}A spinning top-shaped rubble pile}",
      journal = {Science},
         year = 2019,
        month = apr,
       volume = {364},
       number = {6437},
        pages = {268-272},
          doi = {10.1126/science.aav8032},
       adsurl = {https://ui.adsabs.harvard.edu/abs/2019Sci...364..268W}
}

@ARTICLE{Scheeres:2019,
       author = {{Scheeres}, D.~J. and {McMahon}, J.~W. and {French}, A.~S. and {Brack}, D.~N. and {Chesley}, S.~R. and {Farnocchia}, D. and {Takahashi}, Y. and {Leonard}, J.~M. and {Geeraert}, J. and {Page}, B. and {Antreasian}, P. and {Getzandanner}, K. and {Rowlands}, D. and {Mazarico}, E.~M. and {Small}, J. and {Highsmith}, D.~E. and {Moreau}, M. and {Emery}, J.~P. and {Rozitis}, B. and {Hirabayashi}, M. and {S{\'a}nchez}, P. and {van Wal}, S. and {Tricarico}, P. and {Ballouz}, R.-L. and {Johnson}, C.~L. and {Al Asad}, M.~M. and {Susorney}, H.~C.~M. and {Barnouin}, O.~S. and {Daly}, M.~G. and {Seabrook}, J.~A. and {Gaskell}, R.~W. and {Palmer}, E.~E. and {Weirich}, J.~R. and {Walsh}, K.~J. and {Jawin}, E.~R. and {Bierhaus}, E.~B. and {Michel}, P. and {Bottke}, W.~F. and {Nolan}, M.~C. and {Connolly}, H.~C. and {Lauretta}, D.~S. and {Osiris-Rex Team}},
        title = "{The dynamic geophysical environment of (101955) Bennu based on OSIRIS-REx measurements}",
      journal = {Nature Astronomy},
         year = 2019,
        month = mar,
       volume = {3},
        pages = {352-361},
          doi = {10.1038/s41550-019-0721-3},
       adsurl = {https://ui.adsabs.harvard.edu/abs/2019NatAs...3..352S}
}

@ARTICLE{Kiefer2017,
       author = {{Kiefer}, F. and {Lecavelier des {\'E}tangs}, A. and {Vidal-Madjar}, A. and {H{\'e}brard}, G. and {Bourrier}, V. and {Wilson}, P.~A.},
        title = "{Detection of a repeated transit signature in the light curve of the enigma star KIC 8462852: A possible 928-day period}",
      journal = {\aap},
         year = 2017,
        month = dec,
       volume = {608},
          eid = {A132},
        pages = {A132},
          doi = {10.1051/0004-6361/201731306},
archivePrefix = {arXiv},
       eprint = {1709.01732},
 primaryClass = {astro-ph.EP},
       adsurl = {https://ui.adsabs.harvard.edu/abs/2017A&A...608A.132K}
}

@ARTICLE{Metzger2017,
       author = {{Metzger}, Brian D. and {Shen}, Ken J. and {Stone}, Nicholas},
        title = "{Secular dimming of KIC 8462852 following its consumption of a planet}",
      journal = {\mnras},
         year = 2017,
        month = jul,
       volume = {468},
       number = {4},
        pages = {4399-4407},
          doi = {10.1093/mnras/stx823},
archivePrefix = {arXiv},
       eprint = {1612.07332},
 primaryClass = {astro-ph.SR},
       adsurl = {https://ui.adsabs.harvard.edu/abs/2017MNRAS.468.4399M}
}

@ARTICLE{Ballesteros2018,
       author = {{Ballesteros}, Fernando J. and {Arnalte-Mur}, Pablo and {Fernandez-Soto}, Alberto and {Mart{\'\i}nez}, Vicent J.},
        title = "{KIC 8462852: Will the Trojans return in 2021?}",
      journal = {\mnras},
         year = 2018,
        month = jan,
       volume = {473},
       number = {1},
        pages = {L21-L25},
          doi = {10.1093/mnrasl/slx105},
archivePrefix = {arXiv},
       eprint = {1705.08427},
 primaryClass = {astro-ph.EP},
       adsurl = {https://ui.adsabs.harvard.edu/abs/2018MNRAS.473L..21B}
}

@ARTICLE{Wright2016,
       author = {{Wright}, Jason T. and {Cartier}, Kimberly M.~S. and {Zhao}, Ming and {Jontof-Hutter}, Daniel and {Ford}, Eric B.},
        title = "{The Search for Extraterrestrial Civilizations with Large Energy Supplies. IV. The Signatures and Information Content of Transiting Megastructures}",
      journal = {\apj},
         year = 2016,
        month = jan,
       volume = {816},
       number = {1},
          eid = {17},
        pages = {17},
          doi = {10.3847/0004-637X/816/1/17},
archivePrefix = {arXiv},
       eprint = {1510.04606},
 primaryClass = {astro-ph.EP},
       adsurl = {https://ui.adsabs.harvard.edu/abs/2016ApJ...816...17W}
}

@ARTICLE{Pearce2021,
       author = {{Pearce}, Logan A. and {Kraus}, Adam L. and {Dupuy}, Trent J. and {Mann}, Andrew W. and {Huber}, Daniel},
        title = "{Boyajian's Star B: The Co-moving Companion to KIC 8462852 A}",
      journal = {\apj},
         year = 2021,
        month = mar,
       volume = {909},
       number = {2},
          eid = {216},
        pages = {216},
          doi = {10.3847/1538-4357/abdd33},
archivePrefix = {arXiv},
       eprint = {2101.06313},
 primaryClass = {astro-ph.SR},
       adsurl = {https://ui.adsabs.harvard.edu/abs/2021ApJ...909..216P}
}

@INPROCEEDINGS{Ricker2014,
       author = {{Ricker}, George R. and {Winn}, Joshua N. and {Vanderspek}, Roland and {Latham}, David W. and {Bakos}, G{\'a}sp{\'a}r. {\'A}. and {Bean}, Jacob L. and {Berta-Thompson}, Zachory K. and {Brown}, Timothy M. and {Buchhave}, Lars and {Butler}, Nathaniel R. and et al.},
        title = "{Transiting Exoplanet Survey Satellite (TESS)}",
    booktitle = {Space Telescopes and Instrumentation 2014: Optical, Infrared, and Millimeter Wave},
         year = 2014,
       editor = {{Oschmann}, Jr., Jacobus M. and {Clampin}, Mark and {Fazio}, Giovanni G. and {MacEwen}, Howard A.},
       series = {Society of Photo-Optical Instrumentation Engineers (SPIE) Conference Series},
       volume = {9143},
        month = aug,
          eid = {914320},
        pages = {914320},
          doi = {10.1117/12.2063489},
       adsurl = {https://ui.adsabs.harvard.edu/abs/2014SPIE.9143E..20R}
}

@ARTICLE{Thorngren2019,
       author = {{Thorngren}, Daniel P. and {Marley}, Mark S. and {Fortney}, Jonathan J.},
        title = "{An Empirical Mass-Radius Relation for Cool Giant Planets}",
      journal = {Research Notes of the American Astronomical Society},
         year = 2019,
        month = sep,
       volume = {3},
       number = {9},
          eid = {128},
        pages = {128},
          doi = {10.3847/2515-5172/ab4353},
       adsurl = {https://ui.adsabs.harvard.edu/abs/2019RNAAS...3..128T}
}

@ARTICLE{Piro2020,
       author = {{Piro}, Anthony L. and {Vissapragada}, Shreyas},
        title = "{Exploring Whether Super-puffs can be Explained as Ringed Exoplanets}",
      journal = {\aj},
         year = 2020,
        month = apr,
       volume = {159},
       number = {4},
          eid = {131},
        pages = {131},
          doi = {10.3847/1538-3881/ab7192},
archivePrefix = {arXiv},
       eprint = {1911.09673},
 primaryClass = {astro-ph.EP},
       adsurl = {https://ui.adsabs.harvard.edu/abs/2020AJ....159..131P}
}

@ARTICLE{Aizawa2017,
       author = {{Aizawa}, Masataka and {Uehara}, Sho and {Masuda}, Kento and {Kawahara}, Hajime and {Suto}, Yasushi},
        title = "{Toward Detection of Exoplanetary Rings via Transit Photometry: Methodology and a Possible Candidate}",
      journal = {\aj},
         year = 2017,
        month = apr,
       volume = {153},
       number = {4},
          eid = {193},
        pages = {193},
          doi = {10.3847/1538-3881/aa6336},
archivePrefix = {arXiv},
       eprint = {1702.08252},
 primaryClass = {astro-ph.EP},
       adsurl = {https://ui.adsabs.harvard.edu/abs/2017AJ....153..193A}
}

@ARTICLE{Christiansen2025,
       author = {{Christiansen}, Jessie L. and {McElroy}, Douglas L. and {Harbut}, Marcy and {Ciardi}, David R. and {Crane}, Megan and {Good}, John and {Hardegree-Ullman}, Kevin K. and {Kesseli}, Aurora Y. and {Lund}, Michael B. and {Lynn}, Meca and {Muthiar}, Ananda and {Nilsson}, Ricky and {Oluyide}, Toba and {Papin}, Michael and {Rivera}, Amalia and {Swain}, Melanie and {Susemiehl}, Nicholas D. and {Tam}, Raymond and {van Eyken}, Julian and {Beichman}, Charles},
        title = "{The NASA Exoplanet Archive and Exoplanet Follow-up Observing Program: Data, Tools, and Usage}",
      journal = {\psj},
         year = 2025,
        month = aug,
       volume = {6},
       number = {8},
          eid = {186},
        pages = {186},
          doi = {10.3847/PSJ/ade3c2},
archivePrefix = {arXiv},
       eprint = {2506.03299},
 primaryClass = {astro-ph.EP},
       adsurl = {https://ui.adsabs.harvard.edu/abs/2025PSJ.....6..186C}
}

@ARTICLE{Rein2019,
       author = {{Rein}, Edan and {Ofir}, Aviv},
        title = "{Fast and precise light-curve model for transiting exoplanets with rings}",
      journal = {\mnras},
         year = 2019,
        month = nov,
       volume = {490},
       number = {1},
        pages = {1111-1119},
          doi = {10.1093/mnras/stz2556},
archivePrefix = {arXiv},
       eprint = {1910.12099},
 primaryClass = {astro-ph.EP},
       adsurl = {https://ui.adsabs.harvard.edu/abs/2019MNRAS.490.1111R}
}

@ARTICLE{Goldreich1982,
       author = {{Goldreich}, P. and {Tremaine}, S.},
        title = "{The dynamics of planetary rings}",
      journal = {\araa},
         year = 1982,
        month = jan,
       volume = {20},
        pages = {249-283},
          doi = {10.1146/annurev.aa.20.090182.001341},
       adsurl = {https://ui.adsabs.harvard.edu/abs/1982ARA&A..20..249G}
}

@ARTICLE{Cuzzi2010,
       author = {{Cuzzi}, J.~N. and {Burns}, J.~A. and {Charnoz}, S. and {Clark}, R.~N. and {Colwell}, J.~E. and {Dones}, L. and {Esposito}, L.~W. and {Filacchione}, G. and {French}, R.~G. and {Hedman}, M.~M. and {Kempf}, S. and {Marouf}, E.~A. and {Murray}, C.~D. and {Nicholson}, P.~D. and {Porco}, C.~C. and {Schmidt}, J. and {Showalter}, M.~R. and {Spilker}, L.~J. and {Spitale}, J.~N. and {Srama}, R. and {Srem{\v{c}}evi{\'c}}, M. and {Tiscareno}, M.~S. and {Weiss}, J.},
        title = "{An Evolving View of Saturn{\textquoteright}s Dynamic Rings}",
      journal = {Science},
         year = 2010,
        month = mar,
       volume = {327},
       number = {5972},
        pages = {1470-1475},
          doi = {10.1126/science.1179118},
       adsurl = {https://ui.adsabs.harvard.edu/abs/2010Sci...327.1470C}
}

@ARTICLE{Boyajian:2018,
       author = {{Boyajian}, Tabetha. S. and {Alonso}, Roi and {Ammerman}, Alex and {Armstrong}, David and {Asensio Ramos}, A. and {Barkaoui}, K. and {Beatty}, Thomas G. and {Benkhaldoun}, Z. and {Benni}, Paul and {Bentley}, Rory O. and {Berdyugin}, Andrei and {Berdyugina}, Svetlana and {Bergeron}, Serge and {Bieryla}, Allyson and {Blain}, Michaela G. and {Capetillo Blanco}, Alicia and {Bodman}, Eva H.~L. and {Boucher}, Anne and {Bradley}, Mark and {Brincat}, Stephen M. and {Brink}, Thomas G. and {Briol}, John and {Brown}, David J.~A. and {Budaj}, J. and {Burdanov}, A. and {Cale}, B. and {Aznar Carbo}, Miguel and {Castillo Garc{\'\i}a}, R. and {Clark}, Wendy J. and {Clayton}, Geoffrey C. and {Clem}, James L. and {Coker}, Phillip H. and {Cook}, Evan M. and {Copperwheat}, Chris M. and {Curtis}, J.~L. and {Cutri}, R.~M. and {Cseh}, B. and {Cynamon}, C.~H. and {Daniels}, Alex J. and {Davenport}, James R.~A. and {Deeg}, Hans J. and {De Lorenzo}, Roberto and {de Jaeger}, Thomas and {Desrosiers}, Jean-Bruno and {Dolan}, John and {Dowhos}, D.~J. and {Dubois}, Franky and {Durkee}, R. and {Dvorak}, Shawn and {Easley}, Lynn and {Edwards}, N. and {Ellis}, Tyler G. and {Erdelyi}, Emery and {Ertel}, Steve and {Farf{\'a}n}, Rafael. G. and {Farihi}, J. and {Filippenko}, Alexei V. and {Foxell}, Emma and {Gandolfi}, Davide and {Garcia}, Faustino and {Giddens}, F. and {Gillon}, M. and {Gonz{\'a}lez-Carballo}, Juan-Luis and {Gonz{\'a}lez-Fern{\'a}ndez}, C. and {Gonz{\'a}lez Hern{\'a}ndez}, J.~I. and {Graham}, Keith A. and {Greene}, Kenton A. and {Gregorio}, J. and {Hallakoun}, Na'ama and {Hanyecz}, Ott{\'o} and {Harp}, G.~R. and {Henry}, Gregory W. and {Herrero}, E. and {Hildbold}, Caleb F. and {Hinzel}, D. and {Holgado}, G. and {Ign{\'a}cz}, Bernadett and {Ilyin}, Ilya and {Ivanov}, Valentin D. and {Jehin}, E. and {Jermak}, Helen E. and {Johnston}, Steve and {Kafka}, S. and {Kalup}, Csilla and {Kardasis}, Emmanuel and {Kaspi}, Shai and {Kennedy}, Grant M. and {Kiefer}, F. and {Kielty}, C.~L. and {Kessler}, Dennis and {Kiiskinen}, H. and {Killestein}, T.~L. and {King}, Ronald A. and {Kollar}, V. and {Korhonen}, H. and {Kotnik}, C. and {K{\"o}nyves-T{\'o}th}, R{\'e}ka and {Kriskovics}, Levente and {Krumm}, Nathan and {Krushinsky}, Vadim and {Kundra}, E. and {Lachapelle}, Francois-Rene and {LaCourse}, D. and {Lake}, P. and {Lam}, Kristine and {Lamb}, Gavin P. and {Lane}, Dave and {Lau}, Marie Wingyee and {Lewin}, Pablo and {Lintott}, Chris and {Lisse}, Carey and {Logie}, Ludwig and {Longeard}, Nicolas and {Lopez Villanueva}, M. and {Whit Ludington}, E. and {Mainzer}, A. and {Malo}, Lison and {Maloney}, Chris and {Mann}, A. and {Mantero}, A. and {Marengo}, Massimo and {Marchant}, Jon and {Mart{\'\i}nez Gonz{\'a}lez}, M.~J. and {Masiero}, Joseph R. and {Mauerhan}, Jon C. and {McCormac}, James and {McNeely}, Aaron and {Meng}, Huan Y.~A. and {Miller}, Mike and {Molnar}, Lawrence A. and {Morales}, J.~C. and {Morris}, Brett M. and {Muterspaugh}, Matthew W. and {Nespral}, David and {Nugent}, C.~R. and {Nugent}, Katherine M. and {Odasso}, A. and {O'Keeffe}, Derek and {Oksanen}, A. and {O'Meara}, John M. and {Ordasi}, Andr{\'a}s and {Osborn}, Hugh and {Ott}, John J. and {Parks}, J.~R. and {Rodriguez Perez}, Diego and {Petriew}, Vance and {Pickard}, R. and {P{\'a}l}, Andr{\'a}s and {Plavchan}, P. and {Pollacco}, Don and {Pozo Nu{\~n}ez}, F. and {Pozuelos}, F.~J. and {Rau}, Steve and {Redfield}, Seth and {Relles}, Howard and {Ribas}, Ignasi and {Richards}, Jon and {Saario}, Joonas L.~O. and {Safron}, Emily J. and {Sallai}, J. Martin and {S{\'a}rneczky}, Kriszti{\'a}n and {Schaefer}, Bradley E. and {Schumer}, Clea F. and {Schwartzendruber}, Madison and {Siegel}, Michael H. and {Siemion}, Andrew P.~V. and {Simmons}, Brooke D. and {Simon}, Joshua D. and {Sim{\'o}n-D{\'\i}az}, S. and {Sitko}, Michael L. and {Socas-Navarro}, Hector and {S{\'o}dor}, {\'A}. and {Starkey}, Donn and {Steele}, Iain A. and {Stone}, Geoff and {Strassmeier}, Klaus G. and {Street}, R.~A. and {Sullivan}, Tricia and {Suomela}, J. and {Swift}, J.~J. and {Szab{\'o}}, Gyula M. and {Szab{\'o}}, R{\'o}bert and {Szak{\'a}ts}, R{\'o}bert and {Szalai}, Tam{\'a}s and {Tanner}, Angelle M. and {Toledo-Padr{\'o}n}, B. and {Tordai}, Tam{\'a}s and {Triaud}, Amaury H.~M.~J. and {Turner}, Jake D. and {Ulowetz}, Joseph H. and {Urbanik}, Marian and {Vanaverbeke}, Siegfried and {Vanderburg}, Andrew and {Vida}, Kriszti{\'a}n and {Vietje}, Brad P. and {Vink{\'o}}, J{\'o}zsef and {von Braun}, K. and {Waagen}, Elizabeth O. and {Walsh}, Dan and {Watson}, Christopher A.},
        title = "{The First Post-Kepler Brightness Dips of KIC 8462852}",
      journal = {\apjl},
         year = 2018,
        month = jan,
       volume = {853},
       number = {1},
          eid = {L8},
        pages = {L8},
          doi = {10.3847/2041-8213/aaa405},
archivePrefix = {arXiv},
       eprint = {1801.00732},
 primaryClass = {astro-ph.SR},
       adsurl = {https://ui.adsabs.harvard.edu/abs/2018ApJ...853L...8B}
}

@ARTICLE{Bourne:2018,
       author = {{Bourne}, R. and {Gary}, B.~L. and {Plakhov}, A.},
        title = "{Recent photometric monitoring of KIC 8462852, the detection of a potential repeat of the Kepler day 1540 dip and a plausible model}",
      journal = {\mnras},
         year = 2018,
        month = apr,
       volume = {475},
       number = {4},
        pages = {5378-5384},
          doi = {10.1093/mnras/sty097},
archivePrefix = {arXiv},
       eprint = {1711.10612},
 primaryClass = {astro-ph.EP},
       adsurl = {https://ui.adsabs.harvard.edu/abs/2018MNRAS.475.5378B}
}

@ARTICLE{Deeg:2018,
       author = {{Deeg}, Hans J. and {Alonso}, Roi and {Nespral}, David and {Boyajian}, Tabetha S.},
        title = "{Non-grey dimming events of KIC 8462852 from GTC spectrophotometry}",
      journal = {\aap},
         year = 2018,
        month = feb,
       volume = {610},
          eid = {L12},
        pages = {L12},
          doi = {10.1051/0004-6361/201732453},
archivePrefix = {arXiv},
       eprint = {1801.00720},
 primaryClass = {astro-ph.SR},
       adsurl = {https://ui.adsabs.harvard.edu/abs/2018A&A...610L..12D}
}
\bibliographystyle{aasjournalv7}



\end{CJK*}
\end{document}